\documentclass[11pt,a4paper]{article}
\usepackage{fullpage}

\usepackage{cite}
\usepackage{microtype}
\usepackage[T1]{fontenc}
\usepackage{amsmath,amssymb}
\usepackage{color}
\usepackage{hyperref}

\newcommand{\config}{\mathbf{C}}
\newcommand{\polylog}{\mbox{\rm polylog}}
\newcommand{\cA}{{\cal A}}
\newcommand{\cB}{{\cal B}}

\newcommand{\cL}{{\cal L}}

\newcommand{\LOCAL}{\textsf{LOCAL}}
\newcommand{\CONGEST}{\textsf{CONGEST}}
\newcommand{\WF}{\textsf{WAIT-FREE}}
\newcommand{\FSYNC}{\textsf{FSYNC}}
\newcommand{\BC}{\mbox{\sf BC}}
\newcommand{\MAD}{\mbox{\sf MAD}}
\newcommand{\MAV}{\mbox{\sf MAV}}
\newcommand{\bc}{\mbox{\tiny{\sf BC}}}
\newcommand{\NBC}{\mbox{\sf NBC}}
\newcommand{\BPBC}{\mbox{\sf BPBC}}

\newcommand{\ld}{\mbox{\tiny{\sf LD}}}
\newcommand{\LD}{\mbox{\sf LD}}
\newcommand{\WFD}{\mbox{\sf WFD}}
\newcommand{\LCL}{\mbox{\sf LCL}}
\newcommand{\BPLD}{\mbox{\sf BPLD}}
\newcommand{\BPNLD}{\mbox{\sf BPNLD}}
\newcommand{\ldo}{\mbox{\tiny{\sf LDO}}}
\newcommand{\LDO}{\mbox{\sf LDO}}
\newcommand{\NLD}{\mbox{\sf NLD}}
\newcommand{\ULD}{\mbox{\sf ULD}}
\newcommand{\UNLD}{\mbox{\sf UNLD}}
\newcommand{\LCP}{\mbox{\sf LCP}}
\newcommand{\PLS}{\mbox{\sf PLS}}
\newcommand{\ALL}{\mbox{\sf ALL}}

\newcommand{\uld}{\mbox{\tiny{\sf ULD}}}
\newcommand{\dga}{\mbox{\tiny{\sf DGA}}}
\newcommand{\wfd}{\mbox{\tiny{\sf WFD}}}
\newcommand{\mad}{\mbox{\tiny{\sf MAD}}}
\newcommand{\ddh}{\mbox{\sf DH}}

\newtheorem{definition}{Definition}

\begin{document}
%%%%%%%%%%%%%%%%%%%%%%%%%%%%%%%%%%%%%%%%%%%%%

\title{Survey of Distributed Decision\thanks{Both authors received additional support from ANR project DESCARTES, and from Inria project GANG.}~\footnote{This is an updated version of the survey appeared in the Bulletin of the EATCS in 2016 (see~\cite{FeuilloleyF16}).}}

\author{Laurent Feuilloley and Pierre Fraigniaud\\ ~\\
\small Institut de Recherche en Informatique Fondamentale\\
\small CNRS and University Paris Diderot }

\date{ }

\maketitle

\begin{abstract}
We survey the recent distributed computing literature on checking whether a given distributed system configuration satisfies a given boolean predicate, i.e., whether the configuration is legal or illegal w.r.t. that predicate. We consider classical distributed computing environments, including mostly synchronous fault-free network computing (\LOCAL\ and \CONGEST\ models), but also asynchronous crash-prone shared-memory computing (\WF\ model),  and mobile computing (\FSYNC\ model). 
\end{abstract}

\newpage 
\tableofcontents
\newpage 

%%%%%%%%%%%%%%%%%%%%%%%%%%%%%%%%%%%%%%%%%%%%%
\section{Introduction}
%%%%%%%%%%%%%%%%%%%%%%%%%%%%%%%%%%%%%%%%%%%%%

The objective of this note is to survey the recent achievements in the framework of \emph{distributed decision}: the computing entities of a distributed system aim at checking whether the system is in a legal state with respect to some boolean predicate. For instance, in a network, the computing entities may be aiming at checking whether the network satisfies some given graph properties. 

Recall that, in a \emph{construction task}, processes have to collectively compute a valid global state of a distributed system, as a collection of individual states, like, e.g.,  providing each node of a network with a color so that to form a proper coloring of that network. Instead, in a \emph{decision task}, processes have to collectively check whether a given global state of a distributed system is valid or not, like, e.g., checking whether a given coloring of the nodes of a network is proper~\cite{Fraigniaud16}. In general, a typical application of distributed decision is checking the validity of outputs produced by the processes w.r.t. a construction task that they were supposed to solved. This applies to various settings, including randomized algorithms as well as algorithms subject to any kind of faults susceptible to corrupt the memory of the processes. 

The  global verdict on the legality of the system state is  obtained as an aggregate of individual opinions produced by all processes. Typically, each process opinion is a single bit (i.e., \emph{accept} or \emph{reject}) expressing whether the system state looks legal or illegal from the perspective of the process, and the global verdict is the \emph{logical conjunction} of these bits. Note that this mechanisms reflects both decision procedures in which the individual opinions of the processes are  collected by some centralized entity, and decision procedures where any process detecting some inconsistency in the  system raises an alarm and/or launches a recovery procedure, in absence of any central entity. We will also briefly consider less common procedures where each process can send some limited information about its environment in the system, and a central authority gathers the information provided by the processes to forge its verdict about the legality of the whole system state.

The difficulty of distributed decision arises when the processes cannot obtain a global perspective of the system, which is typically the case if one insists on some form of locality in networks, or if the processes are asynchronous and subject to failures. In such frameworks, not all boolean predicates on distributed systems can be checked in a distributed manner, and one of the main issue of distributed decision is to characterize the predicates that can be distributedly checked, and at which cost. For predicates that cannot be checked, or for which checking is too costly, the system can be enhanced by providing processes with \emph{certificates}, with the objective to help these processes for expressing their individual opinions. Such certificates could be produced by an external entity, but they might also well be produced by the processes themselves during a pre-computation phase. One typical framework in which the latter scenario finds application is self-stabilization. Indeed, a self-stabilizing algorithm may produce, together with its distributed output, a distributed certificate that this output is correct. Of course, the certificates are also corruptible, and thus not trustable. Hence, the checking procedure must involve a distributed verification algorithm in charge of verifying the collection of pairs (output, certificate) produced by all the processes. Some even more elaborated mechanisms for checking the legality of distributed system states are considered in the literature, and we survey such mechanisms as well. 

We consider the most classical distributed computing models, including synchronous distributed network computing~\cite{Peleg00}. In this setting, processes are nodes of a graph representing a network. They all execute the same algorithm, they are fault-free, and they are provided with distinct identities in some ID-space (which can be bounded or not). All processes start simultaneously, and computation proceeds in synchronous rounds. At each round, every process exchanges messages with its neighboring processes in the network, and performs individual computation.  The volume of communication each node can transmit and receive on each of its links at each round might be bounded or not. The  \CONGEST\ model typically assumes that at most $O(\log n)$ bits can be transferred along each link at each round in $n$-node networks. (In this case, the ID-space is supposed to be polynomially bounded as a function of the network size).  Instead the  \LOCAL\ model does not limit the amount of information that can be transmitted along each link at each round. So, a $t$-round algorithm $\cA$ in the  \LOCAL\ model can be transformed into another algorithm $\cB$ in which every node first collects all data available in the ball of radius~$t$ around it, and, second, simulate~$\cA$ locally without communication.  

We also consider other models like asynchronous distributed shared-memory computing~\cite{AttiyaW04}.  In this setting, every process has access to a global memory shared by all processes. Every process accesses this memory via atomic read and write instructions. The memory is composed of registers, and each process is allocated a set of private registers. Every process can read all the registers, but can only write in its own registers. Processes are given distinct identities in $[n]=\{1,\dots,n\}$ for $n$-process systems. They runs asynchronously, and are subject to crashes. A process that crashes stops taking steps. An arbitrary large number of processes can crash. Hence, an algorithm must never include instructions leading a process to wait for actions by another process, as the latter process can crash. This model is thus often referred to as the \WF\ model. 

Finally, we briefly consider other models, including mobile computing~\cite{FlocchiniPS12}, mostly in the fully-synchronous \FSYNC\ model in graphs (where all mobile agents perform in lock-step, moving from nodes to adjacent nodes in a network), and distributed quantum computing (where processes have access to intricate variables).

%%%%%%%%%%%%%%%%%%%%%%%%%%%%%%%%%%%%%%%%%%%%%
\section{Model and Definitions}
%%%%%%%%%%%%%%%%%%%%%%%%%%%%%%%%%%%%%%%%%%%%%

Given a boolean predicate, a distributed decision algorithm is a distributed algorithm in which every process $p$ must eventually output a value 
\[
\mbox{opinion}(p)\in\{\mbox{accept, reject}\}
\] 
such that the global system state satisfies the given predicate if and only if all processes accept. In other word, the global interpretation of the individual opinions produced by the processes  is the  logical conjunction of all these opinions:
\[
\mbox{global verdict} = \bigwedge_p \mbox{opinion}(p).
\]
Among the earliest references explicitly related to distributed decision, it is worth mentioning~\cite{AfekKY97,AwerbuchPV91,ItkisL94}. In this section, we describe the general framework of distributed decision, without explicit references to some specific underlying computational model. 
 
The structure of the section is inspired from the structure of complexity classes in sequential complexity theory. Given the ``base'' class $\mathsf{P}$ of languages that are sequentially decidable by a Turing machine in time polynomial in the size of the input, the classes $\mathsf{NP}$ (for non-deterministic polynomial time) and $\mathsf{BPP}$ (for bounded probability polynomial time) are defined, as well as the classes $\Sigma_k^{\tiny\mathsf{P}}$ and $\Pi_k^{\tiny\mathsf{P}}$, $k\geq 0$, of the polynomial hierarchy. 
In this section we assume given an abstract class $\BC$ (for \emph{bounded distributed computing}), based on which larger classes can be defined. Such a base class $\BC$ could be a \emph{complexity} class like, e.g., the class of graph properties that can be checked in constant time in the \LOCAL\ model, or a \emph{computability} class like, e.g., the class of system properties that can be checked in a shared-memory distributed system subject to crash  failures. Given the ``base'' class $\BC$, we shall define the classes $\NBC$, $\BPBC$, $\Sigma_k^{\bc}$ and $\Pi_k^{\bc}$, that are to $\BC$ what $\mathsf{NP}$, $\mathsf{BPP}$, $\Sigma_k^{\tiny\mathsf{P}}$ and $\Pi_k^{\tiny\mathsf{P}}$ are to $\mathsf{P}$, respectively. 

%----------------------------------
\subsection{Distributed Languages}
%----------------------------------

A system \emph{configuration} $\config$ is a (partial) description of a  distributed system state. For instance, in distributed network computing,  a configuration $\config$  is of the form $(G,\ell)$ where $G$ is a graph, and $\ell:V(G)\to \{0,1\}^*$. Similarly, in shared memory computing, a configuration $\config$  is of the form $\ell:[n] \to \{0,1\}^*$ where $n$ is the number of processes. The function $\ell$ is called \emph{labeling} function, and $\ell(v)$ the \emph{label} of~$v$, which can be any arbitrary bit string. In the context of distributed decision, the label of a process is the input of that process. 

For instance, the label of a node in a processor network can be a color, and the label of a process in a shared memory system can be a status like ``elected'' or ``defeated''. Note that, in both examples, a configuration is oblivious to the content of the shared memory and/or to the message in transit. The labeling function $\ell$ may not  describe the full state of each process, but only the content of some specific variables.  

\begin{definition}
Given a distributed computing model, a \emph{distributed language}  is a Turing-computable set of configurations compatible with this model.
\end{definition}

 For instance, in the framework of network computing, 
\[
\mbox{\sc proper-coloring}=\{(G,\ell): \forall \{u,v\}\in E(G), \ell(u)\neq \ell(v)\}
\]
is the distributed language composed of all networks with a proper coloring of their nodes (the label $\ell(v)$ of node~$v$ is its color). Similarly, in the framework of crash-prone shared-memory computing,
\[
\mbox{\sc agreement}=\{\ell :  \exists y\in\{0,1\}^*, \forall i\in [n], \ell(i)=y \;\mbox{or} \; \ell(i)=\bot \}
\]
is the distributed language composed of all systems where agreement between the non-crashed processes is achieved (the label of process $p_i$ is $\ell(i)$, and the symbol~$\bot$ refers to the scenario in which process $p_i$ crashed). 

For a fixed distributed language $\cL$, a configuration in $\cL$ is said to be \emph{legal}, and a configuration not in $\cL$ is said to be \emph{illegal}. Any distributed language $\cL$ defines a \emph{construction} task, in which every process must compute a  label such that the collection of labels outputted by the processes form a legal configuration for $\cL$. In the following, we are mostly interested in \emph{decision} tasks, where the labels of the nodes are given, and the processes must collectively check whether these labels form a legal configuration. 

\paragraph{Notation.}

Given a system configuration $\config$ with respect to some distributed computing model, we denote by $V(\config)$ the set of all computing entities (a.k.a. processes) in $\config$. This notation reflects the fact that, in the following, the set of processes will most often be identified as the vertex-set $V(G)$ of a graph~$G$

%----------------------------------
\subsection{Distributed Decision}
%----------------------------------

Given a distributed computing model, let us define some \emph{bounded computing} class $\BC$ as a class of distributed languages that can be decided with a distributed algorithm~$\cA$ using a bounded amount of resources. Such an algorithm~$\cA$ is said to be \emph{bounded}. What is meant by ``resource'' depends on the computing model. In most of the models investigated in this paper, the resource of interest is the number of rounds (as in  the \LOCAL\ and \CONGEST\ models), or the number of read/write operations (as in the \WF\ model). A distributed language $\cL$ is in $\BC$ if and only if there exists a bounded algorithm $\cA$ such that, for any input configuration $\config$, the algorithm $\cA$ outputs $\cA(\config,v)$ at each process $v$, and this output satisfies: 
\begin{equation}\label{eq:IoA}
\config\in \cL \iff \mbox{for every $v\in V(\config)$,} \; \cA(\config,v)=\mbox{accept}.
\end{equation}
That is, for every $\config\in \cL$, running $\cA$ on $\config$ results in all processes accepting~$\config$. Instead, for every $\config\notin \cL$, running $\cA$ on $\config$ results in  at least one process rejecting~$\config$. 

\paragraph{Example.}

In the context of network computing, {\sc proper-coloring} can be decided in one round, by having each node merely comparing its color with the ones of its neighbors, and accepting if and only if its color is different from all these colors. Similarly, in the context of shared-memory computing, {\sc agreement} can be decided by having each node performing just one read/write operation, accepting if and only if all labels different from~$\bot$ observed in memory  are identical. In other words, assuming that  \BC\/ is a network computing class bounding algorithms to perform in a constant number of rounds, we have
\[
\mbox{\sc proper-coloring} \in \BC
\]
for any model allowing each process to send its color to all its neighbors in a constant number of rounds, like, e.g., the \LOCAL\ model. Similarly, assuming that \BC\/ is a shared-memory computing class bounding algorithms to perform in a constant number of read/write operations, we have
\[
\mbox{\sc agreement} \in \BC.
\]

\paragraph{Notation.}

In the following, Eq.~\eqref{eq:IoA} will  often be abbreviated to
\[
\config\in \cL \iff \cA(\config)=\mbox{accept}
\]
in the sense that $\cA$ accepts if and only if each of the processes accepts. 

\medskip 

Note that the rule of distributed decision, i.e., the logical conjunction of the individual boolean outputs of the processes is not symmetric. For instance, deciding whether a graph is properly colored can be done locally, while deciding whether a graph is \emph{not} properly colored may require long-distance communications. On the other hand, asking for other rules, like unanimous decision (where all processes must reject an illegal configuration) or even just majority decision, would require long-distance communications for most classical decision problems. 

%----------------------------------
\subsection{Probabilistic Distributed Decision}
%----------------------------------

The bounded computing class $\BC$ is a base class upon which other classes can be defined. Given  $p,q\in [0,1]$, we define the class $\BPBC(p,q)$, for \emph{bounded probability bounded computing},  as the class of all distributed languages $\cL$ for which there exists a randomized bounded algorithm $\cA$ such that, for every configuration~$\config$, 
\begin{equation}
\left\{\begin{array}{lcl}
\config \in \cL & \Rightarrow &  \Pr[\cA(\config) =\mbox{accept}]\geq p; \\
\config \notin \cL & \Rightarrow &  \Pr[ \cA(\config) =\mbox{reject}]\geq q.
\end{array}\right. 
\end{equation}
Such an algorithm $\cA$ is called a $(p,q)$-decider for $\cL$. Note that, as opposed to the class \textsf{BPP} of complexity theory, the parameters $p$ and $q$ are not arbitrary, in the sense that boosting the probability of success of a $(p,q)$-decider in order to get a $(p',q')$-decider with $p'>p$ and $q'>q$ is not always possible. Indeed, if $\cA$ is repeated many times on an illegal instance, say $k$ times, it may well be the case that each node will reject at most once during the $k$ repetitions, because, at each iteration of $\cA$, rejection could come from a different node. As a consequence, classical boosting techniques based on repetition and taking majority do not necessarily apply. 

\paragraph{Example.} Let us consider the following distributed language, where each process can be labeled either white or black, i.e., $\ell:V(\config)\to \{\circ,\bullet\}$: 
\[
\mbox{\sc amos}=\{\ell: |\{v\in V(\config) : \ell(v)=\bullet \}|\leq 1 \}. 
\]
Here, {\sc amos} stands for ``at most one selected'', where a node $v$ is selected if $\ell(v)=\bullet$. There is a trivial $(p,q)$-decider for  {\sc amos}  as long as $p^2+q\leq 1$, which works as follows. Every node $v$ with $\ell(v)=\circ$ accepts (with probability~1). A node $v$ with $\ell(v)=\bullet$ accepts with probability~$p$, and rejects with probability $1-p$. If $\config\in \mbox{\sc amos}$, then $\Pr[\mbox{all nodes accept} \; \config]\geq p$. If $\config\notin \mbox{\sc amos}$, then $\Pr[\mbox{at least one node rejects} \; \config]\geq 1-p^2\geq q$. 

%----------------------------------
\subsection{Distributed Verification}
%----------------------------------

Given a bounded computing class $\BC$, we describe the class $\NBC$, which is to $\BC$ what \textsf{NP} is  to \textsf{P} in complexity theory.  We define the class $\NBC$, for \emph{non-deterministic bounded computing},  as the class of all distributed languages $\cL$ such that there exists a bounded algorithm $\cA$ satisfying that, for every configuration~$\config$, 
\begin{equation}\label{eq:nld}
\config \in \cL \iff \exists c : \cA(\config,c) =\mbox{accept}
\end{equation}
where 
\[
c: V(\config)\to \{0,1\}^*.
\]
The function $c$ is called the \emph{certifying} function. It assigns a certificate to every process, and the certificates do not need to be identical. Note that the certificate $c(v)$ of process~$v$ must not be mistaken with the label $\ell(v)$ of that process. 

The bounded algorithm $\cA$ is also known as a \emph{verification} algorithm for $\cL$, as it verifies a given proof $c$, which is supposed to certify that $\config\in \cL$. At each process $v\in V(\config)$, the verification algorithm takes as input the pair $(\ell(v),c(v))$. Note that the appropriate certificate $c$ leading to accept a configuration $\config\in\cL$ may depend on the given configuration $\config$. However, for $\config\notin \cL$, the verification algorithm $\cA$ must systematically guaranty that at least one process rejects, whatever the given certificate function is. 

Alternatively, one can interpret Eq.~\eqref{eq:nld} as a game between a \emph{prover} which, for every configuration $\config$, assigns a certificate $c(v)$  to each process $v\in V(\config)$, and a  \emph{verifier} which checks that the certificates assigned by the prover collectively form a \emph{proof} that $\config\in \cL$. For a legal configuration (i.e., a configuration in $\cL$) the prover must be able to produce a distributed proof leading the distributed verifier to accept, while, for an illegal configuration, the verifier must reject in at least one node whatever the proof provided by the prover is. 

\paragraph{Example.} Let us consider the distributed language 
\[
\mbox{\sc acyclic}=\{(G,\ell): \mbox{$G$ has no cycles} \}
\]
in the context of network computing. Note that $\mbox{\sc acyclic}$ cannot be decided locally, even in the $\LOCAL$ model. However, {\sc acyclic} can be verified in just one round. If $G$ is acyclic, i.e.,  $G$ is a forest, then let us  select an arbitrary node in each tree of $G$, and call it a root. Next, let us assign to each node $u\in V(G)$ the certificate $c(u)$ equal to its distance to the root of its tree. The verification algorithm $\cA$ then proceeds at every node $u$ as follows. Node $u$ exchanges its certificate with the ones of it neighbors, and checks that it has a unique neighbor $v$ satisfying $c(v)=c(u)-1$, and all the other neighbors $w \neq v$ satisfying  $c(w)=c(u)+1$. (If $u$ has $c(u)=0$, then it checks that all its neighbors $w$ have $c(w)=1$). If all tests are passed, then $u$ accepts, else it rejects. If $G$ is a acyclic, then, by construction, the verification accepts at all nodes. Instead, if $G$ has a cycle, then, for every setting of the certifying function, some inconsistency will be detected by at least one node of the cycle, which leads this node to  reject. Hence 
\[
\mbox{\sc acyclic}\in \NBC
\]
where $\BC$ bounds the number of rounds, for every distributed computing model allowing every node to exchange $O(\log n)$ bits along each of its incident edges at every round, like, e.g., the \CONGEST\ model. 

\paragraph{Notation.} 

For any function $f:\mathbb{N}\to \mathbb{N}$, we define $\NBC(f)$ as the class $\NBC$ where the certificates are bounded to be on at most $f(n)$ bits in $n$-node networks. For $f\in \Theta(\log n)$,  $\NBC(f)$ is rather denoted by $\mbox{log-}\NBC$.  

%----------------------------------
\subsection{Distributed Decision Hierarchy}
%----------------------------------

In the same way the polynomial hierarchy \textsf{PH} is built upon \textsf{P} using alternating universal and existential quantifiers, one can define a hierarchy built upon base class  \BC. Given a  class $\BC$ for some distributed computing  model, we define the \emph{distributed decision hierarchy} $\ddh^{\bc}$ as follows. We set $\Sigma_0^{\bc}=\Pi_0^{\bc}=\BC$, and, for $k\geq 1$, we set $\Sigma_k^{\bc}$ as the class of all distributed languages $\cL$ such that there exists a bounded algorithm $\cA$  satisfying that, for every configuration~$\config$, 
\[
\config \in \cL \iff \exists c_1\, \forall c_2 \; \exists c_3 \dots Q c_k : \cA(\config,c_1,\dots,c_k) =\mbox{accept}
\]
where, for every $i\in \{1,\dots,k\}$, $c_i: V(\config)\to \{0,1\}^*$, and $Q$ is the universal quantifier if $k$ is even, and the existential one otherwise. The class $\Pi_k^{\bc}$ is defined similarly, by having a universal quantifier as first quantifier, as opposed to an existential one as in~$\Sigma_k^{\bc}$.  The $c_i$'s are  called \emph{certifying} functions. In particular, we have 
\[
\NBC = \Sigma_1^{\bc}.
\]
Finally, we define 
\[
\ddh^{\bc}= \left (\cup_{k\geq 0} \; \Sigma_k^{\bc} \right ) \cup (\cup_{k\geq 0} \; \Pi_k^{\bc}).
\]
As for $\NBC$, a class $\Sigma_k^{\bc}$ or $\Pi_k^{\bc}$ can be viewed as a game between a prover (playing  the existential quantifiers), a disprover (playing the universal quantifiers), and a verifier (running a verification algorithm $\cA$). 

\paragraph{Example.} Let us consider the distributed language 
\[
\mbox{\sc vertex-cover}=\big\{(G,\ell): \{v\in V(G): \ell(v)=1\}\; \mbox{is a minimum vertex cover}\big  \}
\]
in the context of network computing. We show that $\mbox{\sc  vertex-cover} \in \Pi_2^{\bc}$, that is, there exists a bounded distributed algorithm $\cA$ such that 
\[
(G,\ell) \in \mbox{\sc  vertex-cover}  \iff \forall c_1\, \exists c_2 : \cA(G,\ell,c_1,c_2) =\mbox{accept}
\]
where \BC\/ is any network computing class bounding algorithms to perform in a constant number of rounds. For any configuration $(G,\ell)$, the disprover tries to provide a vertex cover $c_1:V(G)\to\{0,1\}$ of size  smaller than the solution~$\ell$, i.e., $|\{v\in V(G):c_1(v)=1\}|<|\{v\in V(G):\ell(v)=1\}|$. On a legal configuration $(G,\ell)$, the prover then reacts by providing each node~$v$ with a certificates $c_2(v)$ such that the $c_2$-certificates collectively encode a spanning tree (and its proof) aiming at demonstrating that there is an error in $c_1$ (like $c_1$ is actually not smaller than $\ell$, or~$c_1$ is not covering some edge, etc.).  It follows that 
\[
\mbox{\sc  vertex-cover} \in \Pi_2^{\bc}
\]
for any model allowing each process to exchange $O(\log n)$-bits messages with its neighbors in a constant number of rounds, like, e.g., the \CONGEST\ model.  

\paragraph{Notation.} 

Similarly to the class $\NBC$, for any function $f:\mathbb{N}\to \mathbb{N}$, we define $\Sigma_k^{\bc}(f)$ (resp., $\Pi_k^{\bc}(f)$) as the class $\Sigma_k^{\bc}$ (resp., $\Pi_k^{\bc}$) where all certificates are bounded to be on at most $f(n)$ bits in $n$-node networks.  For $f\in \Theta(\log n)$, these classes are denoted by $\mbox{log-}\Sigma_k^{\bc}$ and $\mbox{log-}\Pi_k^{\bc}$, respectively. The classes $\ddh^{\bc}(f)$ and $\mbox{log-}\ddh^{\bc}$ are defined similarly. 

%%%%%%%%%%%%%%%%%%%%%%%%%%%%%%%%%%%%%%%%%%%%%
\section{Distributed Decision in Networks}
%%%%%%%%%%%%%%%%%%%%%%%%%%%%%%%%%%%%%%%%%%%%%

In this section, we focus on languages defined as collections of configurations of the form $(G,\ell)$ where $G$ is a simple connected $n$-node graph, and $\ell:V(G)\to \{0,1\}^*$ is a labeling function assigning to every node~$v$ a label $\ell(v)$. Recall that an algorithm $\cA$ is  \emph{deciding} a distributed language $\cL$ if and only if, for every configuration $(G,\ell)$, 
\[
(G,\ell) \in \cL \iff \cA(G,\ell) \; \mbox{accepts at all nodes.}
\]
 
%----------------------------------
\subsection{\LOCAL\ model}
%----------------------------------

% - - - - - - - - - - - - - - - - - - -
\subsubsection{Local Distributed Decision ($\LD$ and $\BPLD$)}
% - - - - - - - - - - - - - - - - - - -

In their seminal paper~\cite{NaorS95}, Naor and Stockmeyer define the class $\LCL$, for \emph{locally checkable labelings}. Let $\Delta \geq 0$, $k\geq 0$, and  $t\geq 0$, and let $\cB$ be a set of balls of radius at most~$t$ with nodes of degree at most $\Delta$, labeled by labels in $[k]$. Note that $\cB$ is finite. Such a set $\cB$ defines the language $\cL$ consisting of all configurations $(G,\ell)$ where $G$ is a  graph with maximum degree~$\Delta$, and $\ell:V(G)\to [k]$, such that all balls of radius~$t$ in $(G,\ell)$ belong to $\cB$. The set $\cB$ is called the set of \emph{good} balls for~$\cL$. $\LCL$ is the class of languages that can be defined by a set of good balls, for some parameters $\Delta$, $k$, and~$t$. For instance the set of $k$-colored graphs with maximum degree~$\Delta$ is a language in $\LCL$. The good balls of this $\LCL$ language are simply the balls of radius~1 where the center node is labeled with a color different from all the colors of its neighbors. 

A series of results were achieved in~\cite{NaorS95} about $\LCL$ languages. In particular, it is Turing-undecidable whether any given $\cL\in \LCL$ has a construction algorithm running in $O(1)$ rounds in the \LOCAL\ model. Also, \cite{NaorS95} showed that the node IDs play a limited role in the context of $\LCL$ languages. Specifically, \cite{NaorS95} proves that, for every $r\geq 0$, if a language $\cL\in \LCL$ has a $r$-round construction algorithm, then it has also a $r$-round \emph{order invariant} construction algorithm, where an algorithm is order invariant if the \emph{relative order} of the node IDs may play a role, but not the actual \emph{values} of these IDs. The assumption  $\cL\in \LCL$ can actually be discarded, as long as $\cL$ remains defined on constant degree graphs with constant labels. That is, \cite{ArfaouiFIM14} proved that, in constant degree graphs, if a language with constant size labels has a $r$-round construction algorithm, then it has also a $r$-round order invariant construction algorithm. Last but not least, \cite{NaorS95} established that randomization is of little help in the context of $\LCL$ languages. Specifically, \cite{NaorS95} proves that if a language $\cL\in \LCL$ has a randomized Monte-Carlo construction algorithm running in $O(1)$ rounds, then $\cL$  also has  a  deterministic construction algorithm running in $O(1)$ rounds. 

The class $\LD$, for \emph{local decision} was defined in~\cite{FraigniaudKP13} as the class of all distributed languages that can be decided in $O(1)$ rounds in the \LOCAL\ model. The class $\LD$ is the basic class playing the role of $\BC$ in the context of local decision. Hence $\LCL\subseteq \LD$ since the set of good balls of a language in $\LCL$ is, by definition, finite. On the other hand, $\LCL\subset \LD$, where the inclusion is strict since $\LD$ does not restrict the graphs to be of bounded degree, nor the labels to be of bounded size. Given $p,q\in[0,1]$, the class $\BPLD(p,q)$, for \emph{bounded probability local decision}, was defined in~\cite{FraigniaudKP13} as the class of languages for which there is a $(p,q)$-decider running in $O(1)$ rounds in the \LOCAL\ model. For $p^2+q\leq 1$, $\BPLD(p,q)$ is shown to include languages that cannot be even decided deterministically in $o(n)$ rounds. On the other hand,  \cite{FraigniaudKP13}  also establishes a derandomization result, stating that, for $p^2+q>1$, if $\cL\in\BPLD(p,q)$, then $\cL\in\LD$. This results however holds only for languages closed under node deletion, and it is proved in~\cite{FraigniaudGKPP14} that, for any every $c \geq 2$, there exists a language $\cL$ with a $(p,q)$-decider satisfying $p^c + q > 1$ and running in a single round, which cannot be decided deterministically in $o(\!\sqrt{n})$ rounds. On the other hand, \cite{FraigniaudGKPP14}   proves that, for $p^2+q>1$, we have $\BPLD(p,q)=\LD$ for all languages restricted on paths. 

On the negative side, it was proved in \cite{FraigniaudGKPP14}  that boosting the probability of success for decision tasks is not always achievable in the distributed setting, by considering the classes
\[
\BPLD_k=\bigcup_{p^{1+1/k}+q>1}\BPLD(p,q) \;\;\mbox{and}\;\; \BPLD_\infty=\bigcup_{p+q>1}\BPLD(p,q)
\] 
for any $k\geq 1$, and proving that,  for every $k\geq 1$, $\BPLD_k \subset \BPLD_\infty$, and $\BPLD_k\subset \BPLD_{k+1}$, where all inclusions are strict. 

On the positive side, it was proved in \cite{FeuilloleyF15}  that the result in~\cite{NaorS95} regarding the derandomization of construction algorithms can be generalized from $\LCL$ to $\BPLD$. Namely, \cite{FeuilloleyF15}  proves that, for languages on bounded degree graphs and bounded size labels, for every $p>\frac12$ and $q>\frac12$, if  $\cL\in \BPLD(p,q)$  has a randomized Monte-Carlo construction algorithm running in $O(1)$ rounds, then $\cL$ has also a  deterministic construction algorithm running in $O(1)$ rounds. 

% - - - - - - - - - - - - - - - - - - -
\subsubsection{Identity-Oblivious Algorithms ($\LDO$)}
% - - - - - - - - - - - - - - - - - - -

In the \LOCAL\ model, a distributed algorithm is \emph{identity-oblivious},  or simply \emph{ID-oblivious}, if the outputs of the nodes are not impacted by the identities assigned to the nodes. That is, for any two ID-assignments given to the nodes, the output of every node must be identical in both cases. Note that an \emph{identity-oblivious} algorithm may use the IDs of the nodes (e.g., to distinguish them), but the output must be oblivious to these IDs. 

The class $\LDO$, for \emph{local decision oblivious} was defined in~\cite{FraigniaudHK12,FraigniaudGKS13},  as the class of all distributed languages that can be decided in $O(1)$ rounds by an ID-oblivious algorithm in the \LOCAL\ model.  The class $\LDO$ is the basic class playing the role of $\BC$ in the context of ID-oblivious local decision. It is shown in~\cite{FraigniaudHK12} that $\LDO= \LD$ when restricted to languages that are closed under node deletion. However, it is proved  in~\cite{FraigniaudGKS13} that $\LDO\subset \LD$, where the inclusion is strict. In the language $\cL\in \LD\setminus\LDO$ used in~\cite{FraigniaudGKS13} to prove the strict inclusion $\LDO\subset \LD$, each node label includes a Turing machine $M$. Establishing $\cL\in \LD$ makes use of an algorithm simulating $M$ at each node, for a number of rounds equal to the identity of the node.  Establishing $\cL\notin \LDO$ makes use of the fact that an ID-oblivious algorithm can be sequentially simulated, and therefore, if an  ID-oblivious algorithm would allow to decide $\cL$, then by simulation of this algorithm, there would exist a sequential algorithm  for separating the set of Turing machines that halts and output~0 from the set of Turing machines that halts and output~1, which is impossible. 

In~\cite{FraigniaudHK12,FraigniaudHS15}, the power of IDs in local decision is characterized using \emph{oracles}. An oracle is a trustable party with full knowledge of the input, who can provide nodes with information about this input. It is shown in~\cite{FraigniaudHK12} that $\LDO\subseteq \LD \subseteq \LDO^{\mbox{\scriptsize \#node}}$ where \#node is the oracle providing each node with an arbitrary large upper bound on the number of nodes. A \emph{scalar} oracle $f$ returns a list $f(n) = (f_1,\dots,f_n)$ of $n$ values that are assigned arbitrarily to the $n$ nodes in a one-to-one manner. A scalar oracle $f$ is large if, for any set of $k$ nodes, the largest value provided by $f$ to the nodes in this set grows with $k$. \cite{FraigniaudHS15} proved that, for any computable scalar oracle $f$, we have $\LDO^f = \LD^f$ if and only if $f$ is large, where $\LD^f$ (resp., $\LDO^f$) is the class of languages that can be locally decided in $O(1)$ rounds in the \LOCAL\ model by an algorithm (resp., by an ID-oblivious algorithm) which uses the information provided by~$f$ available at the nodes. 

% - - - - - - - - - - - - - - - - - - -
\subsubsection{Anonymous Networks}
% - - - - - - - - - - - - - - - - - - -

Derandomization results were achieved in~\cite{EmekPSW14} in the framework of anonymous network (that is, nodes have no IDs). Namely, for every language $\cL$ that can be decided locally in any anonymous network, if there exists a randomized anonymous construction algorithm for $\cL$, then there exists a deterministic anonymous construction algorithm for $\cL$, provided that the latter is equipped with a 2-hop coloring of the input network. In addition, \cite{EmekSW14}  shows that, in anonymous networks, giving the ability to the nodes of revoking  their decisions (i.e., to change it as long as not all the nodes have output) considerably changes the power of the model. 

%----------------------------------
\subsection{\CONGEST\ model}
%----------------------------------

% - - - - - - - - - - - - - - - - - - -
\subsubsection{Decision Algorithms}
% - - - - - - - - - - - - - - - - - - -

In \cite{KorKP13} and \cite{SarmaHKKNPPW12} the authors consider decision problems such as checking whether a given set of edges forms a spanning tree, checking whether a given set of edges forms a minimum-weight spanning tree (MST), checking various forms of connectivity, etc. All these decision tasks require essentially $\Theta(\sqrt{n} + D)$ rounds (the lower bound is typically obtained using reduction to communication complexity). In particular,  \cite{SarmaHKKNPPW12}  proved that  checking whether a given set of edges is a  spanning tree requires $\Omega(\sqrt{n} + D)$ rounds, which is much more that what is required to construct a spanning tree ($O(D)$ rounds, using a simple breadth-first search). However,  \cite{SarmaHKKNPPW12}  proved that, for some other problems (e.g., MST), lower bounds on the round-complexity of the decision task consisting in checking whether a solution is valid  yield lower bounds on the round-complexity of the corresponding construction task, and this holds also for the construction of approximate solutions. (In \cite{ElkinKNP14}, the techniques of \cite{SarmaHKKNPPW12} are extended to a quantum setting, to achieve similar lower bounds).

% - - - - - - - - - - - - - - - - - - -
\subsubsection{Distributed Property Testing}
% - - - - - - - - - - - - - - - - - - -

Very few distributed languages on graphs can be checked locally in the \CONGEST\ model. For instance, even  just deciding whether $G$ contains a triangle cannot be done in $O(1)$ rounds in the \CONGEST\ model. \emph{Distributed property testing} is a framework first investigated in~\cite{BrakerskiP11}, and recently formalized in~\cite{Censor-HillelFS16}. Let $0<\epsilon <1$ be a fixed parameter. Recall that, according to the usual definition borrowed from \emph{property testing} (in the so-called \emph{sparse} model), a graph property $P$ is $\epsilon$-far from being satisfied by an $m$-edge graph $G$ if applying a sequence of at most $\epsilon m$ edge-deletions or edge-additions to $G$ cannot result in a graph satisfying $P$. We say that a distributed algorithm $\cA$ is a distributed \emph{testing} algorithm for $P$ if and only if, for any graph $G$ modeling the actual network, 
\[
\left \{ \begin{array}{l}
\mbox{$G$ satisfies $P \implies \Pr[\cA \; \mbox{accepts  $G$ in all nodes}]\geq \frac23$;}  \\
\mbox{$G$ is $\epsilon$-far from satisfying $P \implies \Pr[\cA\; \mbox{rejects $G$ in at least one node}]\geq \frac23$.}
\end{array} \right.
\]
Among other results, \cite{Censor-HillelFS16} proved that, in bounded degree graphs, bipartiteness can be distributedly tested in $O(\polylog\, n)$ rounds in the \CONGEST\ model. Moreover, it is also proved that triangle-freeness can be distributedly tested in $O(1)$ rounds. (The dependence in $\epsilon$ is hidden in the big-$O$ notation). This latter result has been recently extended in~\cite{FraigniaudRST16} to testing $H$-freeness, for every 4-node graph $H$,  in $O(1)$ rounds. On the other hand, it is not known whether distributed testing $K_5$-freeness or $C_5$-freeness can be achieved  in $O(1)$ rounds, and~\cite{FraigniaudRST16}  proves that ``natural'' approaches based on DFS or BFS traversals do not work. Testing $C_k$-freeness is $O(1)$ rounds for every $k\geq 3$ has been achieved in \cite{FraigniaudO17}. These results have been recently generalized in~\cite{EvenFF+17}, by proving that $T$-freeness can be decided in $O(1)$ rounds (see also~\cite{KorhonenR17}). Finally, tight bounds on testing graph conductance are provided in~\cite{FichtenbergerV17}. 

% - - - - - - - - - - - - - - - - - - -
\subsubsection{Congested clique} 
% - - - - - - - - - - - - - - - - - - -

The \emph{congested clique} model is the \CONGEST\ model restricted to complete graphs. Deciding whether a graph given as input  contains some specific patterns as subgraphs has been considered in  \cite{Censor-HillelKK15} and \cite{DolevLP12} for the congested clique. In particular, \cite{Censor-HillelKK15} provides an algorithm for deciding the presence of a $k$-node cycle $C_k$ running in  $O(2^{O(k)}n^{0.158})$-rounds. 

Recently, \cite{KorhonenS17} defined the class {\sf CLIQUE}($t$) as the class of decision problems that can be decided in time $t$ in the congested clique, and  provided a collection of separation results regarding these classes, as a function of $t$. 

In \cite{KlauckNP015} an $\Omega(n/k^2)$ lower bound on the number of rounds for deciding properties such as connectivity and spanning tree is provided for the \emph{$k$-machine model}, with matching upper bounds in \cite{Pandurangan0S16}. (In the $k$-machine model, the edges of the input graph are randomly partitioned among $k$ machines that are linked by a clique, and these $k$ machines proceed as in the \CONGEST\/ model with bandwidth limited to one single bit per round). 

%----------------------------------
\subsection{General Interpretation of Individual Outputs}
\label{subsec:giio}
%----------------------------------

In~\cite{ArfaouiFIM14,ArfaouiFP13}, a generalization of distributed decision is considered, where every node output not just a single bit (accept or reject), but can output an arbitrary bit-string. The global verdict is then taken based on the multi-set of all the binary strings outputted by the nodes. The concern is restricted to decision algorithms performing in $O(1)$ rounds in the \LOCAL\ model, and the objective is to minimize the size of the outputs. The corresponding basic class $\BC$ for outputs on $O(1)$ bits is denoted by $\ULD$, for \emph{universal $\LD$}. (It is universal in the sense that the global interpretation of the individual outputs is not restricted to the logical conjunction). It is proved in~\cite{ArfaouiFIM14} that, for any positive even integer $\Delta$, every distributed decision algorithm for cycle-freeness in connected graphs with degree at most $\Delta$ must produce outputs of size at least $\lceil \log \Delta\rceil- 1$ bits. Hence, cycle-freeness does not belong to $\ULD$ in general, but it does belong to $\ULD$ for constant degree graphs. 

In \cite{BeckerKMNRST15} the authors consider a model in which each node initially knows the IDs of its neighbors, while the nodes do not communicate through the edges of the network but via a public whiteboard. The concern of \cite{BeckerKMNRST15}  is mostly restricted to the case in which every node can write only once on the whiteboard, and the objective is to minimize the size of the message written by each node on the whiteboard. The global verdict is then taken based on the collection of messages written on the whiteboard. It is shown that, with just $O(\log n)$-bit messages,  it is possible to rebuild the whole graph from the information on the whiteboard as long as the graph is planar or, more generally, excluding a fixed minor. Variants of the model are also considered, in which problems such as deciding triangle-freeness or connectivity are considered. See also~\cite{KariMRS15} for deciding the presence of induced subgraphs. 

%%%%%%%%%%%%%%%%%%%%%%%%%%%%%%%%%%%%%%%%%%%%%
\section{Distributed Verification in Networks}
%%%%%%%%%%%%%%%%%%%%%%%%%%%%%%%%%%%%%%%%%%%%%

In this section, we still focus on languages defined as collections of configurations of the form $(G,\ell)$ where $G$ is a simple connected $n$-node graph, and $\ell:V(G)\to \{0,1\}^*$ is a labeling function. Recall that an algorithm $\cA$ is  \emph{verifying} a distributed language $\cL$ if and only if, for every configuration $(G,\ell)$, 
\begin{equation}\label{eq:roleID0}
(G,\ell) \in \cL \iff \exists c : \cA(G,\ell,c) \; \mbox{accepts at all nodes}
\end{equation}
where $c:V(G)\to\{0,1\}^*$, and $c(v)$ is called the \emph{certificate} of node $v\in V(G)$. Again, the certificate $c(v)$ of node~$v$ must not be mistaken with the label $\ell(v)$ of node~$v$. Also, the notion of certificate must not be confused with the notion of \emph{advice}. While the latter  are trustable information provided by an oracle~\cite{FraigniaudGIP09}, the former are proofs that must be verified. 

We survey the results about the class $\NBC=\Sigma_1^{\bc}$ where the basic class $\BC$ is $\LD$, $\LDO$, $\ULD$, etc.

%----------------------------------
\subsection{\LOCAL\ model}
%----------------------------------

It is crucial to distinguish two cases in Eq.~\eqref{eq:roleID0}, depending on whether the certificates can depend on the identities assigned to the nodes, or not, as reflected in Eq.~\eqref{eq:roleID1} and~\eqref{eq:roleID2} below. 

% - - - - - - - - - - - - - - - - - - -
\subsubsection{Local Distributed Verification ($\Sigma_1^{\ld}$, $\PLS$, and $\LCP$)}
\label{subsec:sigma1ld}
% - - - - - - - - - - - - - - - - - - -

A distributed language $\cL$ satisfies $\cL\in \Sigma_1^{\ld}$ if and only if there exists a verification algorithm $\cA$ running in $O(1)$ rounds in the \LOCAL\ model such that,  for every configuration $(G,\ell)$, we have 
\begin{equation}\label{eq:roleID1}
\left \{ \begin{array}{lcl}
(G,\ell) \in \cL & \Rightarrow &  \forall  \mbox{ID}, \exists c, \; \cA(G,\ell,c) \; \mbox{accepts at all nodes}\\
(G,\ell) \notin \cL & \Rightarrow &  \forall \mbox{ID}, \forall c, \; \cA(G,\ell,c) \; \mbox{rejects in at least one node}
\end{array}\right .
\end{equation}
where $c:V(G)\to\{0,1\}^*$, and where, for $(G,\ell) \in \cL$, the assignment of the certificates to the nodes may depend on the identities given to these nodes. This notion has actually been introduced under the terminology \emph{proof-labeling scheme} in~\cite{KormanKP10}, where the concern is restricted to verification algorithms running in just a single round, with the objective of minimizing the size of the certificates. In particular, it is proved that minimum-weight spanning tree can be verified with certificates on $O(\log^2 n)$ bits in $n$-node networks, and this bound in tight~\cite{KormanK07} (see also~\cite{KorKP13}). Interestingly, the $\Omega(\log^2n)$ bits lower bound on the certificate size can be broken, and reduced to $O(\log n)$ bits, to the price of allowing verification to proceed in $O(\log^2 n)$ rounds~\cite{KormanKM15}. There are tight connections between proof-labeling schemes and compact silent self-stabilizing algorithms~\cite{BlinFP14}, and proof-labeling schemes can even be used as a basis to semi-automatically derive  compact time-efficient self-stabilizing algorithms~\cite{BlinF15}. Let $\PLS$ be the class of distributed languages for which there exists a proof-labeling scheme. We have 
\[
\PLS=\ALL
\] 
where $\ALL$ is the class of all distributed languages on networks (i.e., with configurations of the form $(G,\ell)$). This equality  is however achieved using certificates on $O(n^2+nk)$ bits in $n$-node networks, where $k$ is the maximum size of the labels in the given configuration $(G,\ell)$. The $O(n^2)$ bits are used to encode the adjacency matrix of the network, and the $O(nk)$ bits are used to encode the inputs to the nodes. 

The notion of proof-labeling scheme has been extended in~\cite{GoosS11} to the notion of \emph{locally checkable proofs}, which is the same as proof-labeling scheme but where the verification algorithm is not bounded to run in a single round, but may perform an arbitrarily large constant number of rounds\footnote{Formally, proof-labeling schemes  assume that the verification algorithm can use only the certificates, and does not have  access to the identifiers of the neighbors, nor to their local information --- this restriction is removed in~\cite{GoosS11}, as it has little impact on the results as long as we are dealing with certificates on $\Omega(\log n)$ bits.}. Let $\LCP$ be the associate class of distributed languages. By definition, we have 
\[
\LCP = \Sigma_1^{\ld},
\]
 and, more specifically,
\[
\LCP(f) = \Sigma_1^{\ld}(f)
\]
for every function $f$ bounding the size of the certificates. Moreover, since $\PLS=\ALL$, it follows that 
\[
\PLS= \LCP= \Sigma_1^{\ld}=\ALL.
\]
It is proved in~\cite{GoosS11} that there are natural languages (e.g., the set of graphs with a non-trivial automorphism, 3-non-colorability, etc.) which require certificates on $\tilde{\Omega}(n^2)$ bits in $n$-node networks. Recently, \cite{BaruchFPS15} introduced a mechanism enabling to reduce exponentially the amount of communication in proof-labeling schemes, using randomization. See also~\cite{SchmidS13} for applications of locally checkable proofs to software-defined networks. 

% - - - - - - - - - - - - - - - - - - -
\subsubsection{Identity-Oblivious Algorithms ($\Sigma_1^{\ldo}$ and $\NLD$)}
\label{subsec:idonld}
% - - - - - - - - - - - - - - - - - - -

A distributed language $\cL$ satisfies $\cL\in \Sigma_1^{\ldo}$ if and only if there exists a verification algorithm $\cA$ running in $O(1)$ rounds in the \LOCAL\ model such that,  for every configuration $(G,\ell)$, we have 
\begin{equation}\label{eq:roleID2}
\left \{ \begin{array}{lcl}
(G,\ell) \in \cL & \Rightarrow &  \exists c, \;  \forall \mbox{ID}, \; \cA(G,\ell,c) \; \mbox{accepts at all nodes}\\
(G,\ell) \notin \cL & \Rightarrow &   \forall c, \; \forall \mbox{ID}, \;  \cA(G,\ell,c) \; \mbox{rejects in at least one node}
\end{array}\right .
\end{equation}
where $c:V(G)\to\{0,1\}^*$, and, for $(G,\ell) \in \cL$, the assignment of the certificates to the nodes must not depend on the identities given to these nodes. In~\cite{FraigniaudKP13}, the class $\NLD$, for  \emph{non-deterministic local decision} is introduced. In $\NLD$, even if the certificates must not depend on the identities of the nodes, the verification algorithm is not necessarily identity-oblivious. Yet, it was proved in~\cite{FraigniaudHK12} that restricting the verification algorithm to be identity-oblivious does not restrict the power of the verifier. Hence, 
\[
\NLD = \Sigma_1^{\ldo}
\]
$\Sigma_1^{\ldo}$ is characterized in~\cite{FraigniaudHK12} as the class of languages that are \emph{closed under lift}, where $H$ is a $k$-lift of $G$ if there exists an homomorphism  from $H$ to $G$ preserving radius-$k$ balls. Hence, 
\[
\Sigma_1^{\ldo}\subset  \ALL
\]
where the inclusion is strict. However, it was proved in~\cite{FraigniaudKP13} that, for every distributed language $\cL$, and for every $p,q$ such that $p^2+q\leq 1$, there is a non-deterministic $(p,q)$-decider for $\cL$.  In other words, for every $p,q$ such that $p^2+q\leq 1$, we have
\[
\BPNLD(p,q)=\ALL.
\]
In~\cite{FraigniaudKP13}, a complete problem for $\NLD$ was identified. However, it was recently noticed in~\cite{BalliuDFO17}  that the notion of local reduction used in~\cite{FraigniaudKP13} is way too strong, enabling to bring languages outside $\NLD$ into $\NLD$. A weaker notion of local reduction was thus defined in~\cite{BalliuDFO17}, preserving the class $\NLD$. A language is  proved to be $\NLD$-complete under this weaker type of local reduction.

%----------------------------------
\subsubsection{Logarithmic-Size Certificates ($\log$-$\Sigma_1^{\ld}$ and $\log$-$\LCP$)}
%----------------------------------

It is shown in~\cite{GoosS11} that many distributed languages can be verified with $\Theta(\log n)$-bit certificates, and hence \cite{GoosS11} investigates the class $\log$-$\LCP$, that is, $\log$-$\Sigma_1^{\ld}$, i.e., $\Sigma_1^{\ld}$ with certificates of size $O(\log n)$ bits. This class fits well with the \CONGEST\ model, which allows to exchange messages of at most $O(\log n)$ bits at each round. For instance, non-bipartiteness is in $\log$-$\LCP$. Also, restricted to bounded-degree graphs,  there are problems in $\log$-$\LCP$ that are not contained in $\mathsf{NP}$, but  $\log$-$\LCP \subseteq \mathsf{NP}/\mathrm{poly}$, i.e., $\mathsf{NP}$ with a polynomial-size non-uniform advice. Last but not least, \cite{GoosS11} shows that existential MSO on connected graphs is included in $\log$-$\LCP$. 

\subsubsection{Approximation} 

The decision languages considered so far are designed to capture decision tasks (e.g., whether a given spanning tree is a MST). For some tasks, such as verifying whether the diameter $D$ of the actual graph is equal to a given value~$k$, there exists no short proof. However, some approximation is easy to certify (e.g., whether $k\leq D \leq 2k$). The notion of \emph{approximate} proof-labeling scheme is defined in~\cite{Censor-HillelPP17}, where the approximate variants of verifying diameter and verifying maximum matching are investigated.

\subsubsection{Space-time trade-offs} 

Except for MST in \cite{KormanKM15}, the running time of the local verification algorithm has always been considered as constant. The impact of allowing larger, non-constant verification times is studied in \cite{OstrovskyPR17}. It is shown that, for several languages, allowing a running time of $t$ yields verification schemes using proofs and message of sizes reduced by a factor $t$ compared to the classical 1-round verification schemes.

\subsubsection{Error-sensitiveness} 

The notion of  \emph{error-sensitive} proof-labeling schemes has been introduced in~\cite{FeuilloleyF17}. Such schemes guarantee that the number of nodes detecting illegal states is linearly proportional to the edit-distance between the current state and the set of legal states. By using  error-sensitive proof-labeling schemes, states which are far from satisfying the predicate will be detected by many nodes, enabling fast return to legality. A structural characterization of the set of boolean predicates on network states for which there exist error-sensitive proof-labeling schemes is provided in~\cite{FeuilloleyF17}. This characterization enables to show that classical predicates such as, e.g., acyclicity, and leader admit error-sensitive  proof-labeling schemes, while others like regular subgraphs don't. Also, it is shown that the known proof-labeling schemes for  spanning tree and minimum spanning tree, using certificates on $O(\log n)$ bits, and on $O(\log^2n)$ bits, respectively, are error-sensitive, as long as the trees are locally represented by adjacency lists, and not just by parent pointers.  

% - - - - - - - - - - - - - - - - - - -
\subsubsection{Anonymous Networks}
% - - - - - - - - - - - - - - - - - - -

Distributed verification in the context of fully anonymous networks (no node-identities, and no port-numbers) has been considered in~\cite{ForsterLSW16}. 

%----------------------------------
\subsection{Message complexity}
%----------------------------------

Some aforementioned papers also pay attention to minimizing the message size. This is explicitly the case of, e.g., \cite{OstrovskyPR17}. 

\subsubsection{Unicast vs. Broadcast.} 

The \CONGEST\/ model has a \emph{broadcast} variant in which each node are restricted to send the same message to all neighbors at each round (the classical \emph{unicast} variant allows each node to send particular messages to each neighbor). The relative powers of unicast and broadcast model in the context of proof-labeling schemes has been studied in~\cite{Patt-ShamirP17}. In particular, it is proved that some languages are insensitive to the broadcast restriction (e.g., spanning tree), whereas others languages (e.g., matching) are significantly impacted by this restriction.

\subsubsection{Congested clique} 

As mentioned before, \cite{KorhonenS17} defined the class {\sf CLIQUE}($t$) as the class of decision problems that can be decided in time $t$ in the congested clique. They also defined the non-deterministic variant of {\sf CLIQUE}($t$), denoted by {\sf NCLIQUE}($t$). A collection of separation results regarding these two classes are provided. The intriguing, and probably challenging open problem of whether $\mbox{\sf CLIQUE}(O(1))=\mbox{\sf NCLIQUE}(O(1))$ is stated in  \cite{KorhonenS17}.

%----------------------------------
\subsection{General Interpretation of Individual Outputs}
\label{subsec:giioderechef}
%----------------------------------

As already mentioned in Section~\ref{subsec:giio}, a generalization of distributed decision was considered in~\cite{ArfaouiFIM14,ArfaouiFP13}, where every node outputs not just a single bit (accept or reject), but can output an arbitrary bit-string. The global verdict is then taken based on the multi-set of all the binary strings outputted by the nodes. The concern is restricted to decision algorithm performing in $O(1)$ rounds in the \LOCAL\ model, and the objective is to minimize the size of the output. The certificates must not depend on the node IDs, that is, verification proceed as specified in Eq.~\eqref{eq:roleID2}. For constant size outputs, it is shown in \cite{ArfaouiFP13} that the class $\UNLD=\Sigma_1^{\uld}$ satisfies 
\[
\UNLD=\ALL
\]
with just 2-bit-per-node outputs, which has to be consider in contrast to the fact that $\NLD$ is restricted to languages that are closed under lift (cf. Section~\ref{subsec:idonld}). This result requires using certificates on $O(n^2+nk)$ bits in $n$-node networks, where $k$ is the maximum size of the labels in the given configuration $(G,\ell)$, but \cite{ArfaouiFP13} shows that this is unavoidable. Also, while verifying cycle-freeness using the logical conjunction of the 1-bit-per-node outputs requires certificates on $\Omega(\log n)$ bits~\cite{GoosS11}, it is proved in~\cite{ArfaouiFP13} that, by simply using the conjunction and the disjunction operators together, on only 2-bit-per-node outputs, one can verify cycle-freeness using certificates of size $O(1)$ bits.

%%%%%%%%%%%%%%%%%%%%%%%%%%%%%%%%%%%%%%%%%%%%%
\section{Local Hierarchies in Networks}
%%%%%%%%%%%%%%%%%%%%%%%%%%%%%%%%%%%%%%%%%%%%%

In this section, we survey the results about the hierarchies $\Sigma_k^{\bc}$ and $\Pi_k^{\bc}$, $k\geq 0$, for different basic classes $\BC$, including $\LD$, $\LDO$, etc. 
%----------------------------------
\subsection{Unlimited-Size Certificates ($\ddh^{\ld}$ and $\ddh^{\ldo}$)}
%----------------------------------

We have seen in Section~\ref{subsec:sigma1ld} that $\Sigma_1^{\ld}=\ALL$, which implies that the local distributed hierarchy $\ddh^{\ld}$ collapses at the first level. On the other hand, we have also seen in  Section~\ref{subsec:idonld} that  $\Sigma_1^{\ldo}\subset \ALL$, where the inclusion is strict as $\Sigma_1^{\ldo}$ is restricted to languages that are closed under lift. It was recently proved in~\cite{BalliuDFO17} that
\[
\LDO \subset \Pi_1^{\ldo} \subset \Sigma_1^{\ldo}  = \Sigma_2^{\ldo} \subset \Pi_2^{\ldo} = \ALL
\]
where all inclusions are strict. Hence, the local ID-oblivious distributed hierarchy collapses at the second level. Moreover, it is shown that $\Pi_2^{\ldo}$ has a complete problem  for local label-preserving reductions. (A complete problem for $\ALL$ was also identified in~\cite{FraigniaudKP13}, but using an inappropriate notion of local reduction). 

In the context of a general interpretation of individual outputs (see Section~\ref{subsec:giioderechef}), \cite{ArfaouiFP13} proved that $\Sigma_1^{\uld}=\ALL$.

%----------------------------------
\subsection{Logarithmic-Size Certificates ($\log$-$\ddh^{\ld}$)}
%----------------------------------

We have previously seen that $\Sigma_1^{\ld}=\ALL$. However, this requires  certificates of polynomial size. The local distributed hierarchy is revisited in~\cite{FeuilloleyFH16}, with certificates of logarithmic size. While it follows from \cite{KormanK07} that $\mbox{\sc mst}\notin \log$-$\Sigma_1^{\ld}$, it is shown in~\cite{FeuilloleyFH16} that 
\[
\mbox{\sc mst}\in \log\!\mbox{-}\Pi_2^{\ld}.
\]
In fact, \cite{FeuilloleyFH16} proved that, for any $k\geq 1$, 
\[
\log\!\mbox{-}\Sigma_{2k}^{\ld}=\log\!\mbox{-}\Sigma_{2k-1}^{\ld} \;\;\mbox{and} \;
\log\!\mbox{-}\Pi_{2k+1}^{\ld}=\log\!\mbox{-}\Pi_{2k}^{\ld},
\]
and thus focused only on the hierarchy $(\Lambda_k)_{k\geq 0}$ defined by $\Lambda_0=\LD$, and, for $k\geq 1$, 
\[
\Lambda_k = \left \{ \begin{array}{ll}
 \log\!\mbox{-}\Sigma_k^{\ld} & \mbox{if $k$ is odd}\\
 \log\!\mbox{-}\Pi_k^{\ld} & \mbox{if $k$ is even}.
\end{array}\right .
\]
It is proved that if there exists $k\geq 0$ such that $\Lambda_{k+1}=\Lambda_k$, then $\Lambda_{k'}=\Lambda_k$ for all $k' \geq k$. That is, the hierarchy collapses at the $k$-th level. Moreover, there exists a distributed language on 0/1-labelled oriented paths that is outside the $\Lambda_k$-hierarchy, and thus outside $\log$-$\ddh^{\ld}$. However, deciding whether a given solution to several optimisation problems such as  maximum independent set, minimum dominating set, maximum matching, max-cut,  min-cut, traveling salesman, etc., is optimal are all in co-$\Lambda_1$, and thus in $\log\!\mbox{-}\Pi_2^{\ld}$. The absence of a non-trivial automorphism is proved to be in $\Lambda_3$, that is $\log\!\mbox{-}\Sigma_3^{\ld}$ --- recall that this language requires certificated of $\tilde{\Omega}(n^2)$ bits to be placed in $\Sigma_1^{\ld}$ (see~\cite{GoosS11}). It is however not known whether $\Lambda_3\neq \Lambda_2$, that is whether $\log\!\mbox{-}\Pi_2^{\ld} \subset \log\!\mbox{-}\Sigma_3^{\ld}$ with a strict inclusion.

% --------------------------------------------
\subsection{Hierarchies in the congested clique}
%  -------------------------------------------

In the congested clique, analogues of the aforementioned hierarchies with unlimited size certificates, as well as with logarithmic size certificates,  are studied in \cite{KorhonenS17}. In particular, it is shown that, as in the \LOCAL\/ model, the hierarchy with  unlimited-size certificates also collapses in the congested clique.

%----------------------------------
\subsection{Distributed Graph Automata  ($\ddh^{\dga}$)}
%----------------------------------

An analogue of the polynomial hierarchy, where sequential polynomial-time computation is replaced by distributed local computation was recently investigated in~\cite{Reiter15}. The model in~\cite{Reiter15} is called \emph{distributed graph automata}. This model assumes a finite-state automaton at each node (instead of a Turing machine), and assumes anonymous computation (instead of the presence of unique node identities). Also, the model assumes an arbitrary interpretation of the outputs produced by each automaton,  based on an arbitrary mapping from the collection of all automata states to $\{\mbox{true},\mbox{false}\}$. The main result in~\cite{Reiter15} is that the hierarchy $\ddh^{\dga}$ coincides with $\mathsf{MSO}$ on graphs.

%%%%%%%%%%%%%%%%%%%%%%%%%%%%%%%%%%%%%%%%%%%%%%
\section{Other Computational Models}
%%%%%%%%%%%%%%%%%%%%%%%%%%%%%%%%%%%%%%%%%%%%%%

%----------------------------------
\subsection{Wait-Free Computing}
%----------------------------------

The class $\WFD$ defined as the class of all distributed languages that are wait-free decidable was characterized in~\cite{FraigniaudRT13} as the class of languages satisfying the so-called  \emph{projection-closeness} property. For non projection-closed languages, \cite{FraigniaudRT14} investigated more general interpretation of the individual opinions produced by the processes, beyond the logical  conjunction of boolean opinons.  In~\cite{FraigniaudRRT14}, it is proved that $k$-set agreement requires that the processes must be allowed to produce essentially $k$ different opinions to be wait-free decided. The class $\Sigma_1^{\wfd}$ has been investigated in~\cite{FraigniaudRT16,FraigniaudRTKR16}, with applications to the space complexity of failure detectors. Interestingly, it is proved in \cite{BonakdarpourFRR16} that wait-free decision finds applications to run-time verification. 

%----------------------------------
\subsection{Mobile Computing}
%----------------------------------

The class $\MAD$, for \emph{mobile agent decision} has been considered in~\cite{FraigniaudP12}, as well as the class $\MAV=\Sigma_1^{\mad}$, for \emph{mobile agent verification}. It is proved that $\MAV$ has a complete language for a basic notion of reduction. The complement classes of $\MAD$ and $\MAV$ have been recently investigated in~\cite{BampasI16} together with sister classes defined by other ways of interpreting the opinions of the mobile agents. 

%----------------------------------
\subsection{Quantum Computing}
%----------------------------------

Distributed decision in a framework in which nodes can have access to extra ressources, such as shared randomness, or intricate variables (in the context of quantum computing) is discussed in~\cite{ArfaouiF14}. In \cite{ElkinKNP14}, the techniques of \cite{SarmaHKKNPPW12} are extended to a quantum setting.

%%%%%%%%%%%%%%%%%%%%%%%%%%%%%%%%%%%%%%%%%%%%%
\section{Conclusion}
%%%%%%%%%%%%%%%%%%%%%%%%%%%%%%%%%%%%%%%%%%%%%

Distributed decision and and distributed verification are known to have applications to very different contexts of distributed computing, including self-stabilization, randomized algorithms, fault-tolerance, runtime verification, etc. In this paper, our aim was to survey the results targeting distributed decision and verification per se. Beside the many interesting problems left open in each of the references listed in this paper, we want to mention two important issues. 

Lower bounds in decision problems are often based on  spatial or temporal arguments. Typically, the  lack of information about far away processes, or the lack of information about desynchronized (or potentially crashed)  processes, prevents processes to forge a consistent opinion about the global status of the distributed system. In the context of shared ressources, such type of arguments appears however to be too weak (cf.~\cite{ArfaouiF14}). Similarly, lower bounds in verification problems are often based on reduction to communication complexity theory. However, such reductions appear to be difficult  to apply to higher classes in the local hierarchy, like separating the class at the third level from the class at the second level of the local hierarchy with $O(\log n)$-bit certificates (cf.~\cite{FeuilloleyFH16}). 

This paper has adopted a systematic approach for presenting the results related to distributed decision and verification from the literature. This approach was inspired from sequential complexity and  sequential computability theories. Such an approach provides a framework that enables to clearly separate decision from verification, as well as clearly separate the results obtained under different assumption (ID-oblivious, size of certificates, etc.). As already mentioned in~\cite{Fraigniaud10}, we believe that distributed decision provides a framework in which bridges between very different models might be identified, as  decision tasks enables easy reductions between languages, while construction tasks are harder to manipulate because of the very different natures of their outputs. 

%%%%%%%%%%%%%%%%%%%%%%%%%%%%%%%%%%%%%%%%%%%%%
\bibliographystyle{plain}
\bibliography{biblio_survey}
%%%%%%%%%%%%%%%%%%%%%%%%%%%%%%%%%%%%%%%%%%%%%
\end{document}